\documentclass[a4paper,fleqn,usenatbib]{mnras}

\usepackage[T1]{fontenc}
\usepackage{ae,aecompl}

\usepackage{graphicx}	
\usepackage{amsmath}	
\usepackage{amssymb}	
\usepackage{hyperref}
\usepackage{pdflscape}	

\title[Ages of galaxies at $z$=0.1-7]
{On the ages of the stellar populations of galaxies at $z$=0.1-7}

\author[Mart\'{\i}nez-Garc\'ia et al.]{
Eric E. Mart\'inez-Garc\'ia $^{1}$\thanks{E-mail: ericmartinez@inaoep.mx}
\\
$^{1}$SECIHTI Research Fellow ~-~Instituto Nacional de Astrof\'isica, \'Optica y Electr\'onica, Luis E. Erro 1, Tonantzintla, Puebla, C.P. 72840, M\'exico\\
}

\date{Accepted XXX. Received YYY; in original form ZZZ}

\pubyear{2025}

\begin{document}
\label{firstpage}
\pagerange{\pageref{firstpage}-~-\pageref{lastpage}}
\maketitle

\begin{abstract}

Recent studies have reported a non evolution of galaxy ages at redshifts
higher than $z\sim$ 2.5, as well as galaxies older than the Universe.
In this work, a sample of galaxies from JWST and HST was
analysed via photometry to further understand this astronomical phenomenon.
No prior cosmological parameters were assumed in the analysis, but the spectroscopic redshift.
When compared to stellar population synthesis models, the results for mass-weighted galaxy ages
indicate that the analysed objects seem to be divided into two subsets.
The results for the subset with the majority of objects (60\% assuming a flat-$\Lambda$CDM cosmology)
indicate an evolution of galaxy ages within the redshift range $z$=0.1-7.0,
in the sense that higher redshift galaxies are younger than the Universe.
Sources of systematic errors were discussed drawing into conclusion that degeneracies between
reddening-age-metallicity, and/or AGN emission may explain the rest 40\% of the galaxies
with ages older than expected from a flat-$\Lambda$CDM cosmology.

\end{abstract}




\section{Introduction}~\label{sec:intro}

Understanding the origin and evolution of the Universe is an important aspect of modern astronomy.
The Big Bang model had led to what is known as the standard cosmology,
aka concordance model, or flat-$\Lambda$CDM model~\citep[e.g.,][]{peeb03}.
The Big Bang model explains most of the observations of the Universe,
for instance,  the discovery of the Hubble flow,
the discovery of the Cosmic microwave background (CMB) radiation,
the prediction of the abundance of light elements, and the baryon acoustic oscillations (BAO) measurements.
However, there are still some observational tests that are taken for granted but have not been fully proven.
For instance, if all galaxies have a common origin,
their stellar populations should be younger than the age of the Universe at a certain redshift.
In this regard, most studies pertain to a limited redshift ($z$) range, and/or a forced restriction
of the ages allowed by the standard cosmological model~\citep[e.g.,][]{mig05,tho17,whi23a,whi23b,dam24}.
There are also studies with no cosmological restrictions
that point to a non evolution of galaxy ages at $z\gtrsim2.5$~\citep{gao24},
or galaxy ages with values greater than those allowed for a flat-$\Lambda$CDM cosmology at $z\sim8$~\citep{lop24}.
In these works, the analysed galaxies present a ``v-shaped'' signature in their spectral energy distributions (SEDs),
where the rest-frame ultraviolet (UV) has a slope $\beta_{\rm UV}\la 0$ and the rest-frame optical slope is $\beta_{\rm opt}>0$.
The ``v-shape'' was interpreted as a mixture of a young stellar population in the UV, and an old stellar population
dominating rest-frame optical wavelengths, giving rise to a Balmer break.

On the other hand, ``Little red dots''~\citep[LRDs, e.g.,][]{mat24,koce24} have been recently uncovered by JWST observations.  
These are distant point-like sources with a ``v-shaped'' SED confirmed by spectroscopy~\citep{fur24,gre24,lab25}.
Stellar light dominating the continuum has been speculated to explain LRDs~\citep[e.g.,][]{lab23,per24,bag24},
with the possibility of having very massive and extremely dense galaxies at early times.
Due to the hosting of broad permitted lines with FWHM$\ga1000$ km/s,
AGN-dominated SEDs have also been proposed~\citep[e.g.,][]{mat24,har23,mai24,koko24,wan24,wan25,ma25,lab25,hai25}.
Recent LRDs developments demonstrate a non-stellar origin for the continuum 
including a Balmer break produced by very dense gas near an
accreting supermassive black hole~\citep{ina25,ji25,deug25,dgraf25,nai25},
which also explains the presence of Balmer absorption lines sometimes detected in LRDs.

The main objective of this work is to further investigate the evolution of galaxy ages
through an analysis unconstrained by any cosmological model, and assess the possible impact of ``v-shape'' objects
in the results.

\section{Observational data}~\label{sec:data}

The observational data consists of two samples of galaxies with data from both the
{\it{Hubble Space Telescope}} (HST) and the~{\it{James Webb Space Telescope}} (JWST).

The photometric data from~\citet{rafel15},
in eleven filters: F225W, F275W, F336W, F435W, F606W, F775W, F850LP, F105W, F125W, F140W and F160W, was adopted for the HST sample.
~\citet{rafel15} computed total AB magnitudes derived from the Kron radius, corrected for Galactic extinction, aperture,
and bandpass variations of the point spread function (PSF).
The spectroscopic redshifts from~\citet{ina17}, with the Multi-Unit
Spectroscopic Explorer~\citep[MUSE,][]{bac10}, were then used to select the objects.
The selection was made based on the confidence level of the redshift.
\citet{ina17}  provides three levels of quality.
Only objects with a confidence level for the redshift of 3 (secure redshift determined by multiple features) were kept.
This gives a total of 263 objects, from which nine of them were discarded since these overlap with the JWST sample (see below).

Data from the JWST Advanced Deep Extragalactic Survey~\citep[JADES,][]{eisen24},
covering the Great Observatories Origins Deep Survey-South~\citep[GOODS-S,][]{will00,cas00}, was adopted for the JWST sample.
Photometric magnitudes in seven wide-band filters from the JWST Near-Infrared
Camera~\citep[NIRCam,][]{riek23a,riek23b}: F090W, F115W, F150W, F200W, F277W, F356W, and F444W;
including also five HST ACS filters: F435W, F606W, F775W, F814W, and F850LP, were utilised.
The photometric magnitudes were obtained with a Kron parameter of 2.5, and corrected for PSF in all HST and JWST filters.
AB magnitudes were obtained and corrected for Galactic extinction using the reddening estimates from~\citet{sch11}.\footnote{This work.}
Objects were selected based on the quality of the spectroscopic redshift adopted from~\citet{deug24},
with the JWST Near-Infrared Spectrograph~\citep[NIRSpec,][]{jak22,fer22}.
~\citet{deug24} assign five $z$ flags. Only objects with flags A or B (highly robust redshifts) were kept.

\begin{figure}
\centering
\includegraphics[width=1.0\hsize]{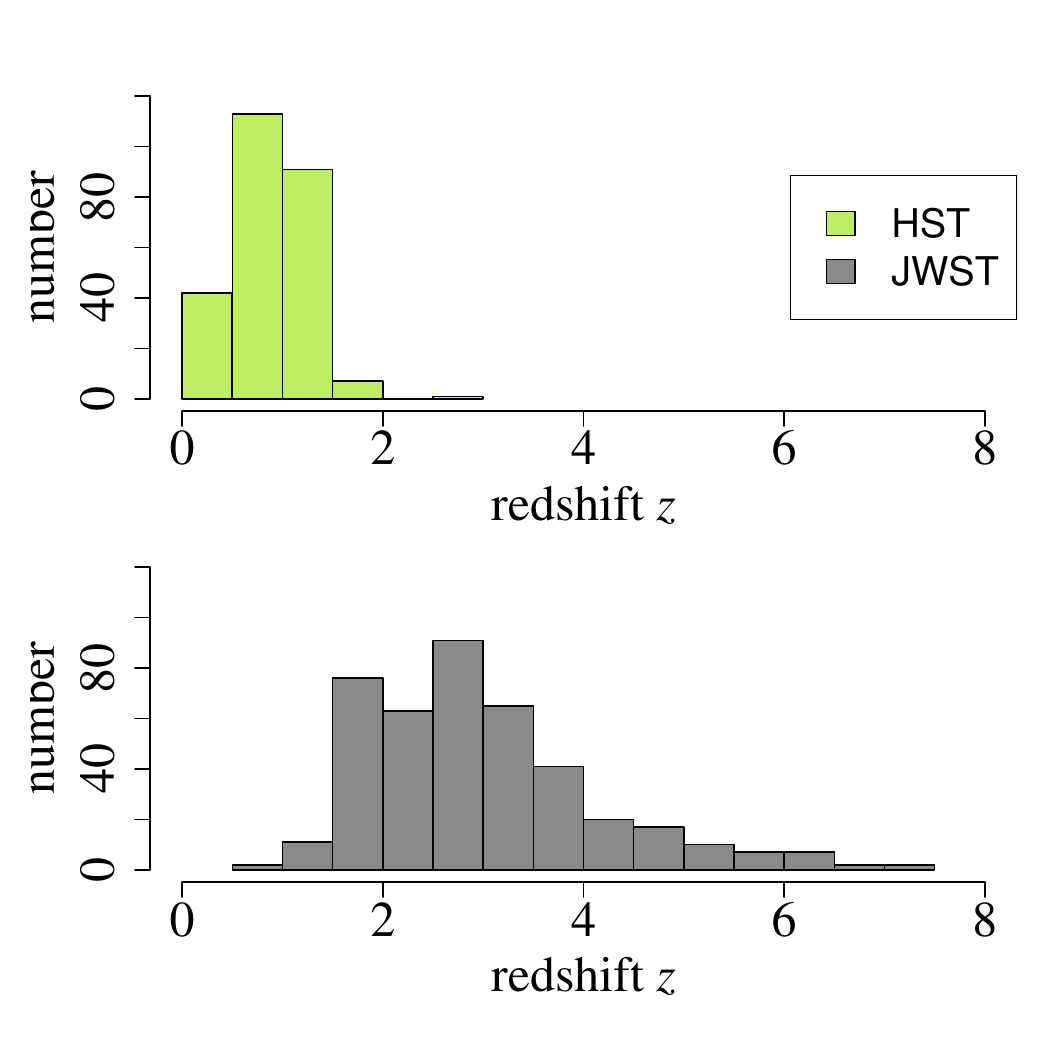}
\caption[fig1]{
Histograms of the spectroscopic redshifts ($z$) of the galaxy sample.
Galaxies with green bins belong to the HST data, while objects with gray bins
to the JWST data.
~\label{fig1}}
\end{figure}

For simplicity reasons the photometry is compared with
models of pure stellar emission (i.e., with no dust emission, see Section~\ref{sec:library}).
The longest rest wavelength ($\lambda$) in NIRcam filters is~$\sim$ 5\micron,
and pure stellar emission stops at $\lambda_{\rm rest}$ of~$\sim$ 2.5\micron~\citep[see, e.g.,][]{daCun08,mart21}.
From here it can be inferred that the lowest redshift for a galaxy to be observed
with JWST NIRcam filters and having only stellar emission is $z\sim1$.
Only 8 objects in the JWST sample have $z\lesssim 1$, these were excluded
from the final sample. The complete sample (HST plus JWST) consists
of 668 objects, 254 from HST and 414 from JWST, respectively.
The redshift range spans from $z=0.127$ to $z=7.433$ (see Figure~\ref{fig1})

\section{Building the SPS libraries}~\label{sec:library}

In order to analyse the photometry of the complete galaxy sample,
libraries of stellar population synthesis (SPS) models were built,
without acknowledging any cosmological parameters, but the redshift.
The SPS models of~\citet[][version 2016]{bc03},\footnote{\url{https://www.bruzual.org/bc03/}.}
with the BaSeL stellar library, and the~\citet{chab03} initial mass
function (IMF, see also Section~\ref{sec:imf}), were used for this purpose.
The libraries were built by randomly selecting the model parameters for the
metallicity, dust attenuation, and star formation history (SFH). 
The main parameter of interest for each model is the mass-weighted age (see Section~\ref{sec:SFH}),
which was used for age dating the objects in the galaxy sample. 
Most of the calculations were done with the with the {\tt{GALAXEV}} software~\citep{bc03}.

\subsection{Metallicity}

The stellar metallicity was distributed according to a 
uniform probability density function (PDF) using six values:
0.0001, 0.0004, 0.004, 0.008, 0.02 ($\sim~Z_{\sun}$), and 0.05.

\subsection{Dust attenuation}

For the dust attenuation, the two component model of~\citet{cha00} was adopted.
In this model the stellar radiation is attenuated by dust in both stellar birth clouds (BC) and the ambient interstellar medium (ISM).
Only dust in the ambient ISM is considered for stellar ages older the typical lifetime of a molecular cloud $\sim1\times10^{7}$~yr.
The absorption optical depth follows the expression $\hat{\tau}_{\lambda}\propto\lambda^{\delta}$,
where $\delta_{\rm{ISM}}=-0.7$ and $\delta_{\rm{BC}}=-1.3$.
The parameters needed to compute the attenuation by dust are the total effective attenuation optical depth
in the $V$ band, $\hat{\tau}_{V}$, and the fraction of $\hat{\tau}_{V}$ originating from the ambient ISM, $\mu$.
The $\hat{\tau}_{V}$ and $\mu$ values were randomly distributed according to the PDFs of~\citet{daCun08}.

\subsection{Star formation history}~\label{sec:SFH}

For the SFH, the delayed model, or~\`a la~\citet{san86},
with an additional burst of star formation~\citep[e.g.,][]{male18,boq19}, was used.
This form of the SFH produces a smooth rise and then a smooth decline for the star formation rate, SFR(t):

\begin{equation}~\label{eq:SFH}
  \Psi(t) = \Psi_{\rm delayed}(t) + \Psi_{\rm burst}(t'),
\end{equation}

\noindent where $\Psi$ in the star formation rate, $t$ is the time since the onset of star formation
with an upper limit of T$_{\rm form}$, $t'=(t-\rm{T}_{\rm form}+\rm{B}_{\rm form})$,
$\rm{B}_{\rm form}$ represents the age of the added star formation burst, and

\begin{eqnarray}
  \Psi_{\rm delayed}(t) & = & (t/\tau_{0}^2) \exp(-t/\tau_{0}),\label{eq:SFH_delayed}\\
  \Psi_{\rm burst}(t')    & = & (k)\exp(-t'/\tau_{1}).\label{eq:SFH_burst}           
\end{eqnarray}

\noindent The parameter $\tau_{0}$ is the e-folding timescale
of the main stellar population and $\tau_{1}$ is the e-folding timescale
of the added burst of star formation.
The factor $k$ is defined as $k\propto\frac{f}{1-f}$~\citep{boq19},
$k=0$ when $t < (\rm{T}_{\rm form}$ - $\rm{B}_{\rm form})$,
where

\begin{equation}
f=\frac{\int_{0}^{(\rm{B}_{\rm form})}\Psi_{\rm burst}(t'){\rm d}t'}
{\int_{0}^{(\rm{T}_{\rm form})}\Psi(t) {\rm d}t},
\end{equation}

\noindent i.e., the fraction of stellar mass formed in the added burst
relative to the total stellar mass ever formed.

In line with the above-mentioned, there are five parameters
that define the SFH for each model: $\tau_{0}$, $\tau_{1}$, $f$, T$_{\rm form}$, and B$_{\rm form}$.
The choice of these parameters will define the mass-weighted age:\\

\begin{equation}~\label{eq:mass-weighted-age}
     {\rm age}_{\rm mw} =
     \frac {\int_{0}^{\rm{T}_{\rm form}} t\Psi(t) {\rm d}t}
           {\int_{0}^{\rm{T}_{\rm form}}  \Psi(t) {\rm d}t},
\end{equation}

\noindent which represents the epoch when most of the stellar mass of a galaxy was assembled~\citep[e.g.,][]{daCun15,lof24},
i.e., an estimate for the age of the object.
A nearly uniform PDF for ${\rm age}_{\rm mw}$, i.e., unrestricted by cosmological parameters, was adopted.
For this purpose, the SFH parameters were randomly selected with the following constraints: 

\begin{itemize} 
   \item $0 < \tau_{0}^{-1} < 1$ (Gyr$^{-1}$),
   \item $10 < \tau_{1} < 100$ (Myr),
   \item $0 < f < 1$,
   \item $100 < {\rm{T}}_{\rm form} < 19990$ (Myr), 
   \item $1 < {\rm{B}}_{\rm form} < {\rm{T}}_{\rm form}$ (Myr).
\end{itemize} 

\noindent The T$_{\rm form}$ parameter upper limit ($\sim20$ Gyr) is given by~\citet{bc03} models.
With each array of parameters ($\tau_{0}$, $\tau_{1}$, $f$, T$_{\rm form}$, and B$_{\rm form}$) we
gradually complete a nearly uniform PDF for ${\rm age}_{\rm mw}$.
This PDF has a smooth decline for ages older than $\ga16$ Gyr, resulting from the fewer parameter
combinations to fill the histogram bins.

\subsection{Redshifting and IGM absorption}~\label{sec:redshifting_IGM}

Given the above conditions, a fiducial SPS library can be constructed. 
During the process it is necessary to rectify each spectrum for the
redshift of the observed object to be analysed,
$z=(\lambda_{\rm observed}-\lambda_{\rm emitted})/\lambda_{\rm emitted}$,
and to correct for the attenuation in the IGM~\citep{mad95}.
The IGM correction was made following the algorithm of~\citet{mei06},
provided in the software ``Code Investigating GALaxy Emission''~\citep[{\tt{CIGALE}},][]{boq19}.

In this manner, for every object in the galaxy sample a library of $\sim$~$4.3\times10^{4}$ models
was built, being the total number of computed models $\sim$~$2.9\times10^{7}$.

\section{Fits to the observed photometry}

The corresponding libraries were used to fit colours of the individual objects in the galaxy sample. 
A maximum likelihood approach was applied by computing the probability:

\begin{equation}~\label{eq:maxlike}
  P \propto \exp \left(-\frac{\chi^{2}}{2}\right),
\end{equation}

\begin{equation}~\label{eq:chi2}
  \chi^{2}=\sum_{i=1}^{n} \left(\frac{C_{i}^{\rm obs}-C_{i}^{\rm model}}{\sigma^{\rm obs}_{i}} \right)^{2},
\end{equation}

\noindent where $C_{i}^{\rm obs}$ is the observed $i_{\rm th}$ colour with $\sigma^{\rm obs}_{i}$ photometric error,
and $C_{i}^{\rm model}$ is the colour of the SPS model.
A total of $n=10$, and $n=11$ colours, with adjacent bandpasses (e.g., F140W-F160W),
were used for the HST and JWST samples, respectively (see Section~\ref{sec:data}).
The use of colours obviates the need to know the distance to the object.
The best fitting model is the one with the maximum probability, $P$, in equation~\ref{eq:maxlike}.
A likelihood distribution for ${\rm age}_{\rm mw}$ can be obtained from the $P$ values,
where the percentiles, P$_{16}$ and P$_{84}$, were computed to estimate the 1-$\sigma$ error.


\section{Results and discussion}~\label{sec:results}

\begin{figure*}
\centering
\includegraphics[width=1.0\hsize]{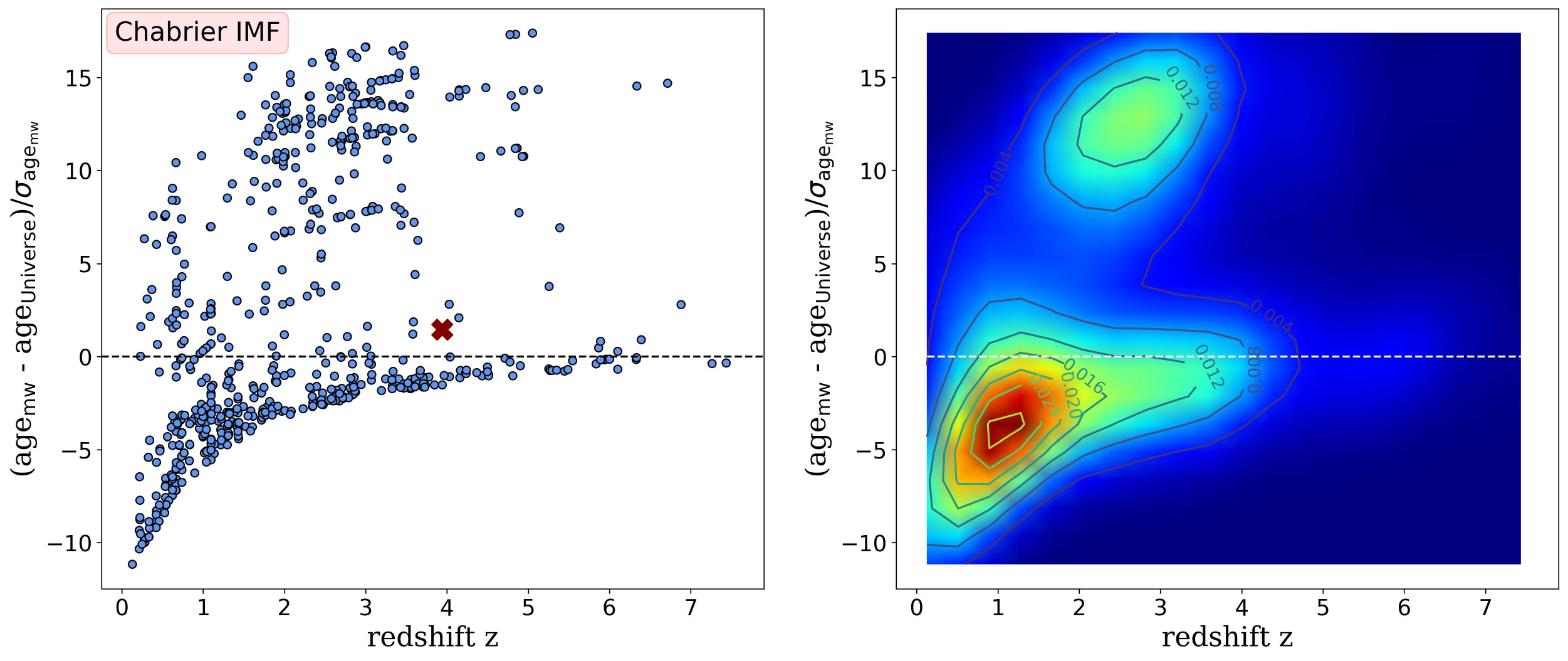}
\caption[fig2]{
{\it Left} panel: $({\rm age}_{\rm mw} - {\rm age}_{\rm Universe})/\sigma_{{\rm age}_{\rm mw}}$
vs. spectroscopic redshift $z$, where ${\rm age}_{\rm mw}$ is the fitted mass-weighted galaxy age 
with random error $\sigma_{{\rm age}_{\rm mw}}$, and ${\rm age}_{\rm Universe}(z)$ is the age of the Universe as a function of redshift
assuming a flat-$\Lambda$CDM cosmology.
The black dashed-line indicates $({\rm age}_{\rm mw} - {\rm age}_{\rm Universe}) = 0$.
The dark red cross shows a LRD source (see Section~\ref{sec:LRDs}).
{\it Right} panel: Density contour plot of the data points in the left panel.
White dashed-line: $({\rm age}_{\rm mw} - {\rm age}_{\rm Universe}) = 0$.
In both panels, results correspond to models with a~\citet{chab03} IMF.
~\label{fig2}}
\end{figure*}

In Figure~\ref{fig2}, left panel, the value of $({\rm age}_{\rm mw} - {\rm age}_{\rm Universe})/\sigma_{{\rm age}_{\rm mw}}$
vs. the spectroscopic redshift $z$, for each object is shown (blue solid points).
The value of ${\rm age}_{\rm Universe}(z)$ was calculated with the cosmology calculator of~\citet{wri06},
with~\citet{ben14} parameters, for a flat-$\Lambda$CDM cosmology.
By analysing the distribution of errors for the complete galaxy sample,
and assuming a Gaussian PDF, a random error of 1-$\sigma_{{\rm age}_{\rm mw}}\sim1.0$ Gyr was estimated.
Systematic errors will be discussed in Section~\ref{sec:systematic_err}.
In Figure~\ref{fig2}, right panel, the density of data points corresponding to the left panel
using contour lines is plotted.
The number of data points with ${\rm age}_{\rm mw}$ younger than ${\rm age}_{\rm Universe}(z)$
is 394 out of 668 ($59\%$), i.e., the majority of sources indicate an evolution of galaxy ages with redshift.
Although the objects seem to be divided into two subsets, where the smaller subset has ages
older than ${\rm age}_{\rm Universe}(z)$. The objects for this subset seem to evolve as older objects for higher redshifts.

\begin{figure}
\centering
\includegraphics[width=1.0\hsize]{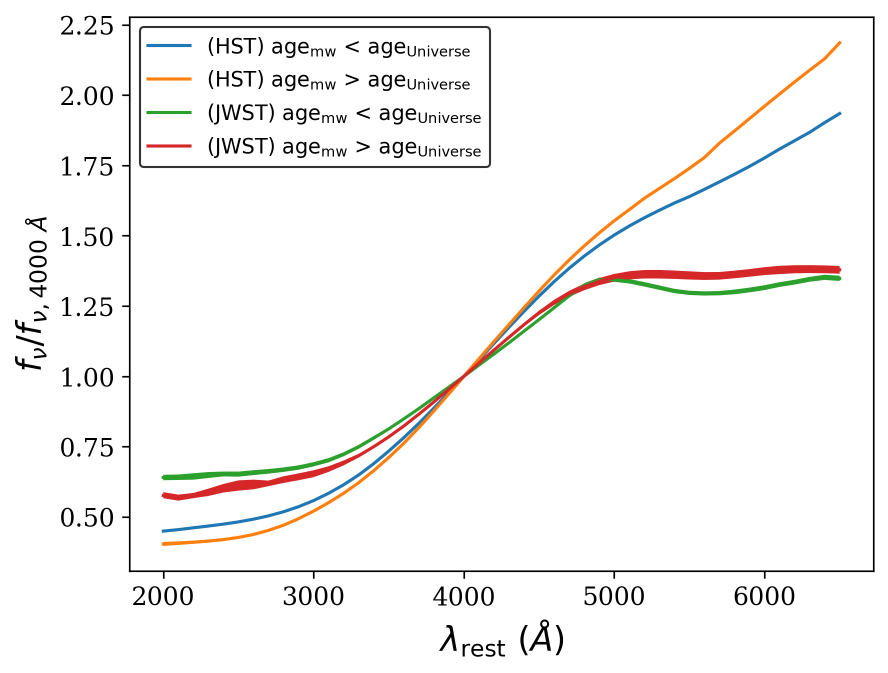}
\caption[fig3]{
Stacked optical rest-frame SEDs.
{\it Blue and orange solid lines}: galaxies from the HST sample.
{\it Green and red solid lines}: galaxies from the JWST sample.
SEDs are grouped according to their retrieved ${\rm age}_{\rm mw}$ with respect to  ${\rm age}_{\rm Universe}(z)$ 
according to the flat-$\Lambda$CDM cosmological model (see Figure~\ref{fig2}).
The width along the $y$ axis of each curve corresponds to an uncertainty of $\sim2\sigma$ of the running mean, obtained via bootstrap methods.
~\label{fig3}}
\end{figure}

\begin{figure}
\centering
\includegraphics[width=1.0\hsize]{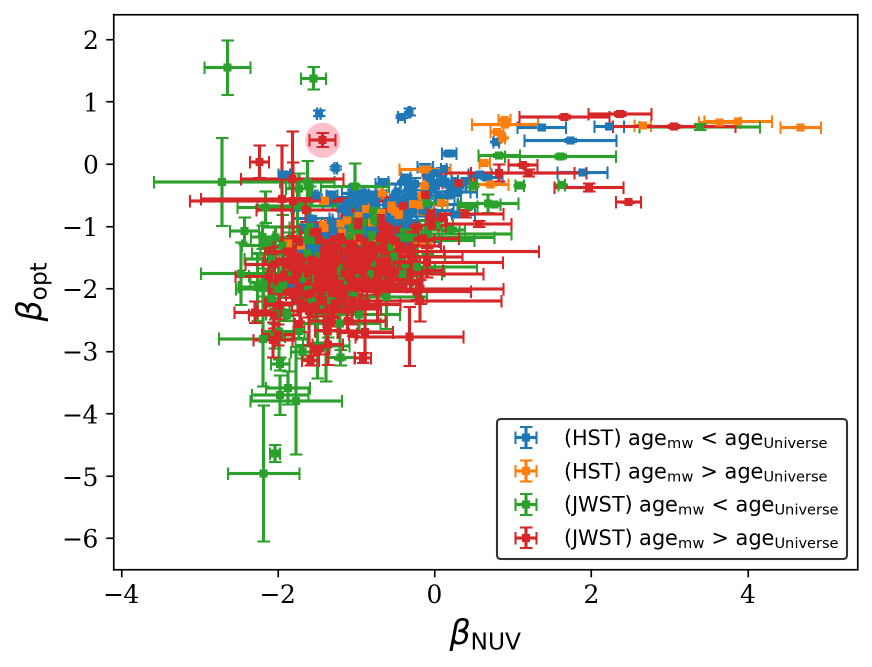}
\caption[fig4]{
Beta slopes, $f_{\lambda}\propto\lambda^{\beta}$.
{\it Blue and orange markers}: HST sample with objects younger and older
than the flat-$\Lambda$CDM Universe, respectively.
{\it Green and red markers}: JWST sample with objects younger and older
than the flat-$\Lambda$CDM Universe, respectively.
The pink solid circle indicates a LRD source (see Section~\ref{sec:LRDs}).
~\label{fig4}}
\end{figure}

In Figure~\ref{fig3}, the stacked optical rest-frame SEDs are shown
for the HST and JWST samples, separately.
The individual SEDs were constructed using the observed photometry\footnote{
For this exercise all the available data in each survey were used, including the F335M and F410M medium-band filters of JADES.}
(see Section~\ref{sec:data}). SEDs were normalised to the flux value at wavelength $\lambda=4000$~\AA~and averaged
within the wavelength range that overlaps for most objects.
The error ($\sigma$) in each average optical SED was computed via bootstrap methods.
In Figure~\ref{fig4}, the individual beta slopes for all the objects in the HST and JWST samples are shown.
The slopes, $\beta_{\mathrm {NUV}}$ and $\beta_{\mathrm {opt}}$, are obtained by fitting a power-law
$f_{\lambda}\propto\lambda^{\beta}$ for the  wavelength intervals 2000(\AA) < $\lambda_{\mathrm {NUV}}$ < 4000(\AA),
and 4000(\AA) < $\lambda_{\mathrm {opt}}$ < 6500(\AA), respectively.
Under the assumption of star-dominated SEDs,
an examination of the plots indicates that the objects in the
JWST sample ($\bar{z}\sim3.0$, green and red markers) have on average younger SEDs as compared to the
objects in the HST sample (mean redshift $\bar{z}\sim0.9$, blue and orange markers),
in the sense that:
$\hat{f}_{\mathrm {NUV}}(\mathrm{HST}) < \hat{f}_{\mathrm {NUV}}(\mathrm{JWST})$, 
and $\hat{f}_{\mathrm {opt}}(\mathrm{HST}) > \hat{f}_{\mathrm {opt}}(\mathrm{JWST})$,
where $\hat{f}_{\mathrm {NUV}}$ and $\hat{f}_{\mathrm {opt}}$ are the normalised
average flux values at the same wavelength intervals as described above.
This qualitative assessment is in accordance with the quantitative
analysis on Figure~\ref{fig2}, indicating an evolution of galaxy ages with redshift.
Also in Figure~\ref{fig3}, the curves have been separated in SEDs with fits indicating age$_{\rm mw}$ < age$_{\rm Universe}$
for a flat-$\Lambda$CDM model (blue and green solid lines) and those where age$_{\rm mw}$ > age$_{\rm Universe}$ (orange and red solid lines).
Again, by assuming star-dominated SEDs, unsurprisingly,
the average optical SEDs for objects with ${\rm age}_{\rm mw}$ values lower than the age of the flat-$\Lambda$CDM Universe
have qualitatively younger SEDs as opposed to the SEDs for galaxies with
ages older than age$_{\rm Universe}$.

Alternatively, if the SEDs are not dominated by stellar emission, then the recovered age$_{\rm mw}$ values
may be biased due to systematic errors as described below.

\subsection{Systematic errors}~\label{sec:systematic_err}

The dispersion of points in Figure~\ref{fig2} suggests that systematic errors 
may deviate some of the age$_{\rm mw}$ results from the true value.
Cosmic variance is assumed to be negligible.
Systematic errors may arise from, e.g., 
degeneracies in the SPS models,
a different stellar IMF, 
emission from thermally pulsing asymptotic giant branch stars,
nebular line and continuum emission, or
sample contamination by ``little red dots'' (and/or AGNs).

\subsubsection{Degeneracies}

Degeneracies between age-metallicity-reddening can give similar colours in the SPS models
causing uncertainty in the results~\citep[see, e.g., ][for $z\sim1$ galaxies]{che25}.
For instance, if degeneracies are not tackled properly
in studies of resolved maps of stellar mass (at $z\sim0$), this may result in biased spatial structures~\citep{mart17}.
This may also have an impact on the retrieved SFH of a galaxy~\citep{mart18}.
Bayesian methods may help to hinder degeneracies, although this kind of
studies are outside the scope of the present investigation.
For the current work, it is assumed that degeneracies can be reduced with a statistically large sample.

\subsubsection{IMF}~\label{sec:imf}

A distinct IMF at different cosmic times may also be a source of systematic error.
It has been suggested that the IMF may evolve with redshift~\citep[e.g.,][]{van08,sne22,ste23}.
The~\citet[][version 2016]{bc03} models provide the alternative to use the~\citet{krou01},~\citet{sal55},
or a top-heavy IMF, besides~\citet{chab03}.
As a means of testing the impact of the IMF choice on the inferred ages,
additional SPS libraries were computed with the~\citet{sal55} IMF, with selfsame parameters
as described in Section~\ref{sec:library}.
The outcome of this exercise results in no significant difference from the~\citet{chab03} IMF case.
In this instance the amount of data points with ages younger than the age of the Universe
is $58\%$, and the trend of younger galaxies at higher redshifts is still verified.

\subsubsection{TP-AGB stars}

Thermally pulsing asymptotic giant branch (TP-AGB) stars are cool luminous giants with stellar masses~$\sim$0.5-6.4~$M_{\sun}$~\citep{mari17}.
Their main contribution to the luminosity of stellar populations lies at rest-frame NIR wavelengths
for ages $\sim 0.2-2$ Gyr~\citep{mara06,bru07}.
The consideration of TP-AGB stars may affect the determination of stellar masses and ages with differences as high as $\sim50\%$.
The inclusion of TP-AGB stars has been favored in some investigations~\citep[e.g.,][]{mara06,mac10,cap16},
and disfavoured in others~\citep[e.g.,][]{kri10,mel12,zib13}.
There is also evidence of a correlation between the presence of TP-AGB stars in nearby disk galaxies and metallicity~\citep[or Hubble type,][]{mart21}. 
For the purpose of testing the effect of the inclusion of TP-AGB stars in the results,
supplementary SPS libraries were constructed.
The~\citet[][version 2016]{bc03} models include a different variation that takes into account
TP-AGB stars with the prescription of~\citet{mari07} and~\citet{mari08},\footnote{\url{https://www.bruzual.org/cb07/}.}
leading to significantly redder NIR colours.
After fitting these libraries to the observed photometry, a similar result to the previous case (Figure~\ref{fig2}) was obtained.
So, although some clear cases of strong TP-AGB signatures have been found in individual galaxies at $z\sim1-2$~\citep{Lu25},
overall the contribution of TP-AGB stars may not be relevant for $z\gtrsim1-2$ galaxies~\citep[see also][]{bev24}.

\subsubsection{Nebular line and continuum emission}

The vast majority of galaxies at $z>3$ are star forming with the implication of
high equivalent widths EW of nebular lines~\citep[100\,\AA\,$\lesssim$ EW $\lesssim$ 1000\,\AA, e.g.,][]{end23,top24}.
Emission lines may bias the results if the flux is boosted in individual filters.
Line-sensitive filters~\citep{rob21} can be used to
test the effect of line boosting of the flux by including in the computations the
JWST medium-band F335M and F410M filters (``M'' filters, which are part of JADES).
In certain cases, these filters trace better the continuum because emission lines will only boost the flux at specific redshifts. 
For this purpose, another set of fits was done by including the JWST ``M'' filters. From these fits the percentage of
objects with ${\rm age}_{\rm mw}$ < ${\rm age}_{\rm Universe}$ is $63\%$,
and $61\%$, for the~\citet{chab03} and~\citet{sal55} IMF's, respectively.
The percentages are slightly higher than the case where no ``M'' filters were adopted (see Figure~\ref{fig2}).
Additionally, continuum nebular emission may become negligible when adopting colours for the fits to the stellar photometry,
as long as this emission remains nearly constant for adjacent filters.
Therefore, for this work it is assumed that nebular emission may bias the results
for some of the objects in the sample but not for the overall outcome.

\subsubsection{LRDs}~\label{sec:LRDs}
  
The isolation of LRDs (and/or AGNs) is necessary to prevent biases in galaxy evolution studies~\citep{chw24}.
Two criteria were adopted as a way to identify possible LRDs in the present JWST sample of galaxies.
The selection technique of~\citet{bar24} involves a colour cut of the form
$(m_{\rm F277W}-m_{\rm F444W}) > 1.5$, where $m_{\rm F277W}$ and $m_{\rm F444W}$ are the magnitudes in the F277W and F444W filters, respectively.
No object in the JWST sample was found to satisfy this colour cut inequality.
The~\citet{gre24} criteria $\texttt{red}\,\texttt{1}$\footnote{Suitable for $4 < z < 6$ objects.}, $\texttt{compact}$,
and $\texttt{v-shape}$ were also applied. Only one object matched the $\texttt{red}\,\texttt{1}$ and $\texttt{compact}$ criteria.\footnote{
The object corresponds to ID 206858 in the JADES NIRCam data~\citep{riek23a}.}
This LRD candidate is highlighted with a dark red cross in Figure~\ref{fig2}.

The results shown in Figures~\ref{fig3} and~\ref{fig4} indicate that the objects where 
${\rm age}_{\rm mw} > {\rm age}_{\rm Universe}(z)$ have on average redder rest-frame $\beta_{\mathrm {opt}}$,
which may indicate a reddened AGN origin for their SEDs, similar to LRDs.
However, the rest-frame $\beta_{\mathrm {NUV}}$ slopes of the same objects do not appear to be bluer as expected for ``v-shaped'' objects.
A detailed spectroscopic analysis would be needed to analyse a possible AGN contribution to the SEDs of the objects studied in this work.


\section{Conclusions}~\label{sec:conclusions}

Based on a sample of 668 galaxies with photometric data from HST and JWST,
spectroscopic redshifts from MUSE and NIRSpec, 
and SPS models that assume no prior cosmological model,
the results demonstrate that for $\sim60\%$ of the analysed objects
the mass-weighted galaxy ages evolve with cosmological redshift, 
in accordance to a flat-$\Lambda$CDM cosmology.
The results hold within the redshift range $0.1\lesssim z \lesssim 7$.
Possible reasons for retrieving $\sim40\%$ of galaxies with ages older than the age of the Universe
may include: degeneracies between age-metallicity-reddening of the SPS models used for the fits,
a different stellar IMF at different redshifts, emission from TP-AGB stars
in the stellar populations, nebular line and continuum emission,
or intruders in the sample due to the presence of LRDs and/or AGNs.
It was discussed that among these, only degeneracies or AGN emission may have a significant effect on the outcome for $z\lesssim7$ galaxies.
Future efforts should focus on eliminating all sources of systematic errors in order to better constrain
galaxy evolution based on observations of galaxy ages.

\section*{Acknowledgements}
The author (EMG) acknowledges the anonymous referee for constructive comments that significantly improved the manuscript.
EMG acknowledges support through the ``Investigadoras e Investigadores por México'' (formerly Cátedras CONACYT) program,
which is now supported by the Secretaría de Ciencia, Humanidades, Tecnología e Innovación (SECIHTI), in México.
EMG acknowledges the use of the {\it Mixtli} server at INAOE from SECIHTI grant number 320772,
administrated by Manuel Zamora-Aviles and Raúl Naranjo (CVU 318178).

\section*{Data availability}

\noindent Based on data obtained from the JWST Advanced Deep Extragalactic Survey (JADES),
\url{https://archive.stsci.edu/hlsp/jades}.





\begin{thebibliography}{}

\bibitem[\protect\citeauthoryear{Bacon et al.}{2010}]{bac10} Bacon R., Accardo M., Adjali L., Anwand H., Bauer S., Biswas I., Blaizot J., et al., 2010, SPIE, 7735, 773508. doi:10.1117/12.856027
\bibitem[\protect\citeauthoryear{Baggen et al.}{2024}]{bag24} Baggen J.~F.~W., van Dokkum P., Brammer G., de Graaff A., Franx M., Greene J., Labb{\'e} I., et al., 2024, ApJL, 977, L13. doi:10.3847/2041-8213/ad90b8
\bibitem[\protect\citeauthoryear{Barro et al.}{2024}]{bar24} Barro G., P{\'e}rez-Gonz{\'a}lez P.~G., Kocevski D.~D., McGrath E.~J., Trump J.~R., Simons R.~C., Somerville R.~S., et al., 2024, ApJ, 963, 128. doi:10.3847/1538-4357/ad167e
\bibitem[\protect\citeauthoryear{Bennett et al.}{2014}]{ben14} Bennett C.~L., Larson D., Weiland J.~L., Hinshaw G., 2014, ApJ, 794, 135. doi:10.1088/0004-637X/794/2/135
\bibitem[\protect\citeauthoryear{Bevacqua et al.}{2025}]{bev24} Bevacqua D., Saracco P., La Barbera F., De Marchi G., De Propris R., Ditrani F., Gallazzi A.~R., et al., 2025, arXiv, arXiv:2501.07291. doi:10.48550/arXiv.2501.07291
\bibitem[\protect\citeauthoryear{Boquien et al.}{2019}]{boq19} Boquien M., Burgarella D., Roehlly Y., Buat V., Ciesla L., Corre D., Inoue A.~K., et al., 2019, A\&A, 622, A103. doi:10.1051/0004-6361/201834156
\bibitem[\protect\citeauthoryear{Bruzual \& Charlot}{2003}]{bc03} Bruzual G., Charlot S., 2003, MNRAS, 344, 1000. doi:10.1046/j.1365-8711.2003.06897.x
\bibitem[\protect\citeauthoryear{Bruzual}{2007}]{bru07} Bruzual A.~G., 2007, IAUS, 241, 125. doi:10.1017/S1743921307007624
\bibitem[\protect\citeauthoryear{Capozzi et al.}{2016}]{cap16} Capozzi D., Maraston C., Daddi E., Renzini A., Strazzullo V., Gobat R., 2016, MNRAS, 456, 790. doi:10.1093/mnras/stv2692
\bibitem[\protect\citeauthoryear{Casertano et al.}{2000}]{cas00} Casertano S., de Mello D., Dickinson M., Ferguson H.~C., Fruchter A.~S., Gonzalez-Lopezlira R.~A., Heyer I., et al., 2000, AJ, 120, 2747. doi:10.1086/316851
\bibitem[\protect\citeauthoryear{Chabrier}{2003}]{chab03} Chabrier G., 2003, PASP, 115, 763. doi:10.1086/376392
\bibitem[\protect\citeauthoryear{Cheng et al.}{2025}]{che25} Cheng C.~M., Kriek M., Beverage A.~G., Slob M., Bezanson R., Franx M., Leja J., et al., 2025, MNRAS, 540, 1527. doi:10.1093/mnras/staf806
\bibitem[\protect\citeauthoryear{Charlot \& Fall}{2000}]{cha00} Charlot S., Fall S.~M., 2000, ApJ, 539, 718. doi:10.1086/309250
\bibitem[\protect\citeauthoryear{Chworowsky et al.}{2024}]{chw24} Chworowsky K., Finkelstein S.~L., Boylan-Kolchin M., McGrath E.~J., Iyer K.~G., Papovich C., Dickinson M., et al., 2024, AJ, 168, 113. doi:10.3847/1538-3881/ad57c1
\bibitem[\protect\citeauthoryear{da Cunha, Charlot, \& Elbaz}{2008}]{daCun08} da Cunha E., Charlot S., Elbaz D., 2008, MNRAS, 388, 1595. doi:10.1111/j.1365-2966.2008.13535.x
\bibitem[\protect\citeauthoryear{da Cunha et al.}{2015}]{daCun15} da Cunha E., Walter F., Smail I.~R., Swinbank A.~M., Simpson J.~M., Decarli R., Hodge J.~A., et al., 2015, ApJ, 806, 110. doi:10.1088/0004-637X/806/1/110
\bibitem[\protect\citeauthoryear{Damjanov, Geller, \& Sohn}{2024}]{dam24} Damjanov I., Geller M.~J., Sohn J., 2024, arXiv, arXiv:2408.05263. doi:10.48550/arXiv.2408.05263
\bibitem[\protect\citeauthoryear{D'Eugenio et al.}{2024}]{deug24} D'Eugenio F., Cameron A.~J., Scholtz J., Carniani S., Willott C.~J., Curtis-Lake E., Bunker A.~J., et al., 2024, arXiv, arXiv:2404.06531. doi:10.48550/arXiv.2404.06531
\bibitem[\protect\citeauthoryear{D'Eugenio et al.}{2025}]{deug25} D'Eugenio F., Maiolino R., Perna M., Uebler H., Ji X., McClymont W., Koudmani S., et al., 2025, arXiv, arXiv:2503.11752. doi:10.48550/arXiv.2503.11752
\bibitem[\protect\citeauthoryear{de Graaff et al.}{2025}]{dgraf25} de Graaff A., Rix H.-W., Naidu R.~P., Labbe I., Wang B., Leja J., Matthee J., et al., 2025, arXiv, arXiv:2503.16600. doi:10.48550/arXiv.2503.16600
\bibitem[\protect\citeauthoryear{Eisenstein et al.}{2023}]{eisen24} Eisenstein D.~J., Willott C., Alberts S., Arribas S., Bonaventura N., Bunker A.~J., Cameron A.~J., et al., 2023, arXiv, arXiv:2306.02465. doi:10.48550/arXiv.2306.02465
\bibitem[\protect\citeauthoryear{Endsley et al.}{2023}]{end23} Endsley R., Stark D.~P., Whitler L., Topping M.~W., Chen Z., Plat A., Chisholm J., et al., 2023, MNRAS, 524, 2312. doi:10.1093/mnras/stad1919
\bibitem[\protect\citeauthoryear{Ferruit et al.}{2022}]{fer22} Ferruit P., Jakobsen P., Giardino G., Rawle T., Alves de Oliveira C., Arribas S., Beck T.~L., et al., 2022, A\&A, 661, A81. doi:10.1051/0004-6361/202142673
\bibitem[\protect\citeauthoryear{Furtak et al.}{2024}]{fur24} Furtak L.~J., Labb{\'e} I., Zitrin A., Greene J.~E., Dayal P., Chemerynska I., Kokorev V., et al., 2024, Natur, 628, 57. doi:10.1038/s41586-024-07184-8
\bibitem[\protect\citeauthoryear{Gao, L{\'o}pez-Corredoira, \& Wei}{2024}]{gao24} Gao C.-Y., L{\'o}pez-Corredoira M., Wei J.-J., 2024, ApJ, 970, 142. doi:10.3847/1538-4357/ad5ce4
\bibitem[\protect\citeauthoryear{Greene et al.}{2024}]{gre24} Greene J.~E., Labbe I., Goulding A.~D., Furtak L.~J., Chemerynska I., Kokorev V., Dayal P., et al., 2024, ApJ, 964, 39. doi:10.3847/1538-4357/ad1e5f
\bibitem[\protect\citeauthoryear{Hainline et al.}{2025}]{hai25} Hainline K.~N., Maiolino R., Juod{\v{z}}balis I., Scholtz J., {\"U}bler H., D'Eugenio F., Helton J.~M., et al., 2025, ApJ, 979, 138. doi:10.3847/1538-4357/ad9920
\bibitem[\protect\citeauthoryear{Harikane et al.}{2023}]{har23} Harikane Y., Zhang Y., Nakajima K., Ouchi M., Isobe Y., Ono Y., Hatano S., et al., 2023, ApJ, 959, 39. doi:10.3847/1538-4357/ad029e
\bibitem[\protect\citeauthoryear{Inami et al.}{2017}]{ina17} Inami H., Bacon R., Brinchmann J., Richard J., Contini T., Conseil S., Hamer S., et al., 2017, A\&A, 608, A2. doi:10.1051/0004-6361/201731195
\bibitem[\protect\citeauthoryear{Inayoshi \& Maiolino}{2025}]{ina25} Inayoshi K., Maiolino R., 2025, ApJL, 980, L27. doi:10.3847/2041-8213/adaebd
\bibitem[\protect\citeauthoryear{Jakobsen et al.}{2022}]{jak22} Jakobsen P., Ferruit P., Alves de Oliveira C., Arribas S., Bagnasco G., Barho R., Beck T.~L., et al., 2022, A\&A, 661, A80. doi:10.1051/0004-6361/202142663
\bibitem[\protect\citeauthoryear{Ji et al.}{2025}]{ji25} Ji X., Maiolino R., {\"U}bler H., Scholtz J., D'Eugenio F., Sun F., Perna M., et al., 2025, arXiv, arXiv:2501.13082. doi:10.48550/arXiv.2501.13082
\bibitem[\protect\citeauthoryear{Kocevski et al.}{2024}]{koce24} Kocevski D.~D., Finkelstein S.~L., Barro G., Taylor A.~J., Calabr{\`o} A., Laloux B., Buchner J., et al., 2024, arXiv, arXiv:2404.03576. doi:10.48550/arXiv.2404.03576
\bibitem[\protect\citeauthoryear{Kokorev et al.}{2024}]{koko24} Kokorev V., Caputi K.~I., Greene J.~E., Dayal P., Trebitsch M., Cutler S.~E., Fujimoto S., et al., 2024, ApJ, 968, 38. doi:10.3847/1538-4357/ad4265
\bibitem[\protect\citeauthoryear{Kriek et al.}{2010}]{kri10} Kriek M., Labb{\'e} I., Conroy C., Whitaker K.~E., van Dokkum P.~G., Brammer G.~B., Franx M., et al., 2010, ApJL, 722, L64. doi:10.1088/2041-8205/722/1/L64
\bibitem[\protect\citeauthoryear{Kroupa}{2001}]{krou01} Kroupa P., 2001, MNRAS, 322, 231. doi:10.1046/j.1365-8711.2001.04022.x
\bibitem[\protect\citeauthoryear{Labb{\'e} et al.}{2023}]{lab23} Labb{\'e} I., van Dokkum P., Nelson E., Bezanson R., Suess K.~A., Leja J., Brammer G., et al., 2023, Natur, 616, 266. doi:10.1038/s41586-023-05786-2
\bibitem[\protect\citeauthoryear{Labbe et al.}{2025}]{lab25} Labbe I., Greene J.~E., Bezanson R., Fujimoto S., Furtak L.~J., Goulding A.~D., Matthee J., et al., 2025, ApJ, 978, 92. doi:10.3847/1538-4357/ad3551
\bibitem[\protect\citeauthoryear{L{\'o}pez-Corredoira et al.}{2024}]{lop24} L{\'o}pez-Corredoira M., Melia F., Wei J.-J., Gao C.-Y., 2024, ApJ, 970, 63. doi:10.3847/1538-4357/ad4f86
\bibitem[\protect\citeauthoryear{Lofaro et al.}{2024}]{lof24} Lofaro C.~M., Rodighiero G., Enia A., Werle A., Bisigello L., Cassata P., Casasola V., et al., 2024, A\&A, 686, A124. doi:10.1051/0004-6361/202347626
\bibitem[\protect\citeauthoryear{Lu et al.}{2025}]{Lu25} Lu S., Daddi E., Maraston C., Dickinson M., Haro P.~A., Gobat R., Renzini A., et al., 2025, NatAs, 9, 128. doi:10.1038/s41550-024-02391-9
\bibitem[\protect\citeauthoryear{Ma et al.}{2025}]{ma25} Ma Y., Greene J.~E., Setton D.~J., Volonteri M., Leja J., Wang B., Bezanson R., et al., 2025, ApJ, 981, 191. doi:10.3847/1538-4357/ada613
\bibitem[\protect\citeauthoryear{MacArthur et al.}{2010}]{mac10} MacArthur L.~A., McDonald M., Courteau S., Jes{\'u}s Gonz{\'a}lez J., 2010, ApJ, 718, 768. doi:10.1088/0004-637X/718/2/768
\bibitem[\protect\citeauthoryear{Madau}{1995}]{mad95} Madau P., 1995, ApJ, 441, 18. doi:10.1086/175332
\bibitem[\protect\citeauthoryear{Maiolino et al.}{2024}]{mai24} Maiolino R., Scholtz J., Curtis-Lake E., Carniani S., Baker W., de Graaff A., Tacchella S., et al., 2024, A\&A, 691, A145. doi:10.1051/0004-6361/202347640
\bibitem[\protect\citeauthoryear{Ma{\l}ek et al.}{2018}]{male18} Ma{\l}ek K., Buat V., Roehlly Y., Burgarella D., Hurley P.~D., Shirley R., Duncan K., et al., 2018, A\&A, 620, A50. doi:10.1051/0004-6361/201833131
\bibitem[\protect\citeauthoryear{Marigo \& Girardi}{2007}]{mari07} Marigo P., Girardi L., 2007, A\&A, 469, 239. doi:10.1051/0004-6361:20066772
\bibitem[\protect\citeauthoryear{Marigo et al.}{2008}]{mari08} Marigo P., Girardi L., Bressan A., Groenewegen M.~A.~T., Silva L., Granato G.~L., 2008, A\&A, 482, 883. doi:10.1051/0004-6361:20078467
\bibitem[\protect\citeauthoryear{Marigo et al.}{2017}]{mari17} Marigo P., Girardi L., Bressan A., Rosenfield P., Aringer B., Chen Y., Dussin M., et al., 2017, ApJ, 835, 77. doi:10.3847/1538-4357/835/1/77
\bibitem[\protect\citeauthoryear{Maraston et al.}{2006}]{mara06} Maraston C., Daddi E., Renzini A., Cimatti A., Dickinson M., Papovich C., Pasquali A., et al., 2006, ApJ, 652, 85. doi:10.1086/508143
\bibitem[\protect\citeauthoryear{Mart{\'\i}nez-Garc{\'\i}a et al.}{2017}]{mart17} Mart{\'\i}nez-Garc{\'\i}a E.~E., Gonz{\'a}lez-L{\'o}pezlira R.~A., Magris C.~G., Bruzual A.~G., 2017, ApJ, 835, 93. doi:10.3847/1538-4357/835/1/93
\bibitem[\protect\citeauthoryear{Mart{\'\i}nez-Garc{\'\i}a et al.}{2018}]{mart18} Mart{\'\i}nez-Garc{\'\i}a E.~E., Bruzual G., Magris C.~G., Gonz{\'a}lez-L{\'o}pezlira R.~A., 2018, MNRAS, 474, 1862. doi:10.1093/mnras/stx2801
\bibitem[\protect\citeauthoryear{Mart{\'\i}nez-Garc{\'\i}a et al.}{2021}]{mart21} Mart{\'\i}nez-Garc{\'\i}a E.~E., Bruzual G., Gonz{\'a}lez-L{\'o}pezlira R.~A., Rodr{\'\i}guez-Merino L.~H., 2021, ApJ, 908, 110. doi:10.3847/1538-4357/abce68
\bibitem[\protect\citeauthoryear{Matthee et al.}{2024}]{mat24} Matthee J., Naidu R.~P., Brammer G., Chisholm J., Eilers A.-C., Goulding A., Greene J., et al., 2024, ApJ, 963, 129. doi:10.3847/1538-4357/ad2345
\bibitem[\protect\citeauthoryear{Meiksin}{2006}]{mei06} Meiksin A., 2006, MNRAS, 365, 807. doi:10.1111/j.1365-2966.2005.09756.x
\bibitem[\protect\citeauthoryear{Melbourne et al.}{2012}]{mel12} Melbourne J., Williams B.~F., Dalcanton J.~J., Rosenfield P., Girardi L., Marigo P., Weisz D., et al., 2012, ApJ, 748, 47. doi:10.1088/0004-637X/748/1/47
\bibitem[\protect\citeauthoryear{Mignoli et al.}{2005}]{mig05} Mignoli M., Cimatti A., Zamorani G., Pozzetti L., Daddi E., Renzini A., Broadhurst T., et al., 2005, A\&A, 437, 883. doi:10.1051/0004-6361:20042434
\bibitem[\protect\citeauthoryear{Naidu et al.}{2025}]{nai25} Naidu R.~P., Matthee J., Katz H., de Graaff A., Oesch P., Smith A., Greene J.~E., et al., 2025, arXiv, arXiv:2503.16596. doi:10.48550/arXiv.2503.16596
\bibitem[\protect\citeauthoryear{Peebles \& Ratra}{2003}]{peeb03} Peebles P.~J., Ratra B., 2003, RvMP, 75, 559. doi:10.1103/RevModPhys.75.559
\bibitem[\protect\citeauthoryear{P{\'e}rez-Gonz{\'a}lez et al.}{2024}]{per24} P{\'e}rez-Gonz{\'a}lez P.~G., Barro G., Rieke G.~H., Lyu J., Rieke M., Alberts S., Williams C.~C., et al., 2024, ApJ, 968, 4. doi:10.3847/1538-4357/ad38bb
\bibitem[\protect\citeauthoryear{Rafelski et al.}{2015}]{rafel15} Rafelski M., Teplitz H.~I., Gardner J.~P., Coe D., Bond N.~A., Koekemoer A.~M., Grogin N., et al., 2015, AJ, 150, 31. doi:10.1088/0004-6256/150/1/31
\bibitem[\protect\citeauthoryear{Rieke et al.}{2023a}]{riek23a} Rieke M.~J., Kelly D.~M., Misselt K., Stansberry J., Boyer M., Beatty T., Egami E., et al., 2023, PASP, 135, 028001. doi:10.1088/1538-3873/acac53
\bibitem[\protect\citeauthoryear{Rieke et al.}{2023b}]{riek23b} Rieke M.~J., Robertson B., Tacchella S., Hainline K., Johnson B.~D., Hausen R., Ji Z., et al., 2023, ApJS, 269, 16. doi:10.3847/1538-4365/acf44d
\bibitem[\protect\citeauthoryear{Roberts-Borsani et al.}{2021}]{rob21} Roberts-Borsani G., Treu T., Mason C., Schmidt K.~B., Jones T., Fontana A., 2021, ApJ, 910, 86. doi:10.3847/1538-4357/abe45b
\bibitem[\protect\citeauthoryear{Salpeter}{1955}]{sal55} Salpeter E.~E., 1955, ApJ, 121, 161. doi:10.1086/145971
\bibitem[\protect\citeauthoryear{Sandage}{1986}]{san86} Sandage A., 1986, A\&A, 161, 89
\bibitem[\protect\citeauthoryear{Schlafly \& Finkbeiner}{2011}]{sch11} Schlafly E.~F., Finkbeiner D.~P., 2011, ApJ, 737, 103. doi:10.1088/0004-637X/737/2/103
\bibitem[\protect\citeauthoryear{Sneppen et al.}{2022}]{sne22} Sneppen A., Steinhardt C.~L., Hensley H., Jermyn A.~S., Mostafa B., Weaver J.~R., 2022, ApJ, 931, 57. doi:10.3847/1538-4357/ac695e
\bibitem[\protect\citeauthoryear{Steinhardt et al.}{2023}]{ste23} Steinhardt C.~L., Kokorev V., Rusakov V., Garcia E., Sneppen A., 2023, ApJL, 951, L40. doi:10.3847/2041-8213/acdef6
\bibitem[\protect\citeauthoryear{Topping et al.}{2024}]{top24} Topping M.~W., Stark D.~P., Endsley R., Whitler L., Hainline K., Johnson B.~D., Robertson B., et al., 2024, MNRAS, 529, 4087. doi:10.1093/mnras/stae800
\bibitem[\protect\citeauthoryear{Thomas et al.}{2017}]{tho17} Thomas R., Le F{\`e}vre O., Scodeggio M., Cassata P., Garilli B., Le Brun V., Lemaux B.~C., et al., 2017, A\&A, 602, A35. doi:10.1051/0004-6361/201628141
\bibitem[\protect\citeauthoryear{van Dokkum}{2008}]{van08} van Dokkum P.~G., 2008, ApJ, 674, 29. doi:10.1086/525014
\bibitem[\protect\citeauthoryear{Wang et al.}{2024}]{wan24} Wang B., Leja J., de Graaff A., Brammer G.~B., Weibel A., van Dokkum P., Baggen J.~F.~W., et al., 2024, ApJL, 969, L13. doi:10.3847/2041-8213/ad55f7
\bibitem[\protect\citeauthoryear{Wang et al.}{2025}]{wan25} Wang B., de Graaff A., Davies R.~L., Greene J.~E., Leja J., Brammer G.~B., Goulding A.~D., et al., 2025, ApJ, 984, 121. doi:10.3847/1538-4357/adc1ca
\bibitem[\protect\citeauthoryear{Whitler et al.}{2023a}]{whi23a} Whitler L., Endsley R., Stark D.~P., Topping M., Chen Z., Charlot S., 2023a, MNRAS, 519, 157. doi:10.1093/mnras/stac3535
\bibitem[\protect\citeauthoryear{Whitler et al.}{2023b}]{whi23b} Whitler L., Stark D.~P., Endsley R., Leja J., Charlot S., Chevallard J., 2023b, MNRAS, 519, 5859. doi:10.1093/mnras/stad004
\bibitem[\protect\citeauthoryear{Williams et al.}{2000}]{will00} Williams R.~E., Baum S., Bergeron L.~E., Bernstein N., Blacker B.~S., Boyle B.~J., Brown T.~M., et al., 2000, AJ, 120, 2735. doi:10.1086/316854
\bibitem[\protect\citeauthoryear{Wright}{2006}]{wri06} Wright E.~L., 2006, PASP, 118, 1711. doi:10.1086/510102
\bibitem[\protect\citeauthoryear{Zibetti et al.}{2013}]{zib13} Zibetti S., Gallazzi A., Charlot S., Pierini D., Pasquali A., 2013, MNRAS, 428, 1479. doi:10.1093/mnras/sts126

\end{thebibliography}






\bsp	
\label{lastpage}
\end{document}